# Observing muon decays in water Cherenkov detectors at the Pierre Auger Observatory

P. Allison, F. Arneodo, X. Bertou, N. Busca, P.L. Ghia, C. Medina, G. Navarra, L. Nellen
H. Salazar Ibarguen, S. Ranchon, M. Urban, L. Villasenor
*for the Pierre Auger Collaboration.*
Presenter: N. Busca (ngb@uchicago.edu) usa-busca-N-abs1-he15-poster

Muons decaying in the water volume of a Cherenkov detector of the Pierre Auger Observatory provide a useful calibration point at low energy. Using the digitized waveform continuously recorded by the electronics of each tank, we have devised a simple method to extract the charge spectrum of the Michel electrons, whose typical signal is about 1/8 of a crossing vertical muon. This procedure, moreover, allows continuous monitoring of the detector operation and of its water level. We have checked the procedure with high statistics on a test tank at the Observatory base and applied with success on the whole array.

## 1. Introduction.

As it is well known, a muon decays into an electron and two neutrinos. The resulting energy distribution of the electron is known as the Michel spectrum. It has an endpoint at 53 MeV and an average of 37 MeV. The range of such an electron is less than 25 cm, whereas the typical dimensions of an Auger tank are of the order of 100 cm. This means that most of the muons decaying in an Auger tank will give rise to electrons well contained inside the tank volume. Because particle type mean energy and mean range are known, the Michel electrons are a calibrated source of Cherenkov light that can be used to measure and monitor a low energy calibration point for the water Cherenkov detectors. Moreover, since the range is a few times smaller than the size of an Auger tank, the signal deposited by a decaying muon will be, to first order, independent of the water level and proportional to the energy of the electron (for very low water level, the electron would not deposit all its energy in the tank and the proportionality would be lost). When this signal is compared to the signal of a crossing muon, which is dependent on water level, a useful observable that allows water level to be monitored is obtained.

The surface detector of the Pierre Auger Observatory is composed of 1600 Cherenkov detector stations (over 800 deployed as of June, 2005). Each station is a cylindrical tank of 3.6 m in diameter and 1.2 m in height, filled with purified and deionized water. The inner surface of the tank is coated with Tyvek, a diffusive-reflective material. Three PMTs mounted on top collect light emitted inside the tank. When the signal satisfies certain trigger requirements, it is digitized by means of a flash ADC (FADC), and stored in a buffer. Signal sizes are expressed in units of $I_{VEM}^{peak}$ for FADC counts, and $Q_{VEM}$ for integrated FADC counts [1]. One $I_{VEM}^{peak}$ is defined as the most likely signal maximum (in FADC counts) given by a vertical crossing muon whereas one $Q_{VEM}$ is defined as the most likely integrated signal (in integrated FADC counts) given by a vertical crossing muon. The conversion factors from integrated FADC counts to $I_{VEM}^{peak}$ and $Q_{VEM}$ are calculated continuously using data from the local station, providing a good measure of the time stability of the signal.

The muon decays in Auger tanks have been the subject of previous works: in [2] the possibility of using a single muon decay as a calibration point was indicated. In this work, a tank smaller than a production Auger tank was used. In [3] a test tank at the Observatory site was purposely depleted of water and its gain was raised to observe Michel electrons more clearly. In [4] a prototype tank in the AGASA site was used to extract muon decay candidates. In [5] a detailed study of the possibility of monitoring the water level by observing muon decays is presented, and two different methods for analyzing data are compared.

This paper is organized as follows. In section 2 we briefly describe the method to distinguish crossing from decaying muon and to obtain the corresponding charge spectra. These data are compared with simulations from which a calibration point is extracted. In section 3 we describe the method for using these spectra to monitor water level and describe the drain experiment carried out on an Auger tank. Finally, in section 4 we discuss our results and conclusions.

## 2. Distinguishing Crossing Muons from Decaying Muons.

Muons entering an Auger water tank can either pass through the tank and exit ("crossing muons") or decay inside the tank ("decaying muons"). The rate of crossing muons is about 2.7 Khz, and each of these crossing muons leaves a signal that is, typically, proportional to the track length inside the tank. Decaying muons, on the other hand, leave a signal proportional to the track length, plus a second pulse corresponding to the Michel electron. These muons are atmospheric muons falling in a water tank as isolated particles, as opposed to muons from high energy showers where a large number of them would cross the tank simultaneously.

To perform this study, we used the "slow buffer" process. The "slow buffer" is a low priority process running in the local station. It collects data from the FADC at a $0.1\ I_{VEM}^{peak}$ threefold (i.e. in the three PMTs) threshold. For each event satisfying this threshold, a 20 bin trace is stored together with a time stamp in a twofold buffer (to minimize dead time). The duration of each of the 20 bins is 25 ns. The time stamp indicates the time at which the trigger condition was satisfied, and corresponds to the third bin of the FADC trace.

Another process in the local station, called "mufill", uses these low threshold data to build three histograms. The first of these histograms is a histogram of the time interval between a given time stamp and the previous one. That is, every time there is a signal satisfying the trigger condition (and hence a corresponding FADC trace in the slow buffer) the time elapsed between this trigger and the previous one is computed using the time stamp of the corresponding traces. A time interval corresponding to a decaying muon trigger followed by a Michel electron trigger will have typical value equal to the muon lifetime. It follows that the resulting histogram will exhibit an exponential decay corresponding to decaying muons over a background corresponding to crossing muons. The time constant of the former should be consistent with the muon lifetime of $2.19\ \mu s$ modified by inverse beta decay capture to about $2.10\ \mu s$ [2].

The second histogram is the histogram of the integrated FADC trace of the "early" events, which are those events that have a time stamp that differs from the previous one in at least $1\ \mu s$ and at most $3\ \mu s$. This first time window, centered at about one muon lifetime, should pick out most of the decaying muons. These events are then candidate Michel electrons from muon decays.

The third histogram is the histogram of the integrated FADC trace of the "late" events, which are those that have a time stamp that differs from the previous one in at least $8.5\ \mu s$ and at most $11\ \mu s$. This time window is centered at over 4 times the lifetime of the muon. These events are then candidate crossing muons.

Figures 1.a to 1.c show these three histograms. From an exponential fit to figure 1.a, we obtain a decay time of $(2.16 \pm 0.05)\ \mu s$ compatible with the decay time mentioned above. The histograms displayed in figures 1.b and 1.c have a similar structure: two peaks at low and high charge. The first peak is produced by the combination of the threefold trigger requirement and low charge noise, consisting of clipping corner crossing muons (i.e. muons with a short track in the tank) as well as electrons from muon decays, and photons from low energy showers.

The second peak is entirely due to crossing muons, most of which cross in directions close to the vertical (the difference of the position of this second peak and the integrated signal deposited by a vertical muon is of the order of $5\%$ [1]). This peak is used for calibration: to obtain the conversion from integrated FADC counts to $Q_{VEM}$, we perform a gaussian fit to the peak. The result of this fit is $1Q_{VEM} = (176 \pm 1)FADC$ counts.

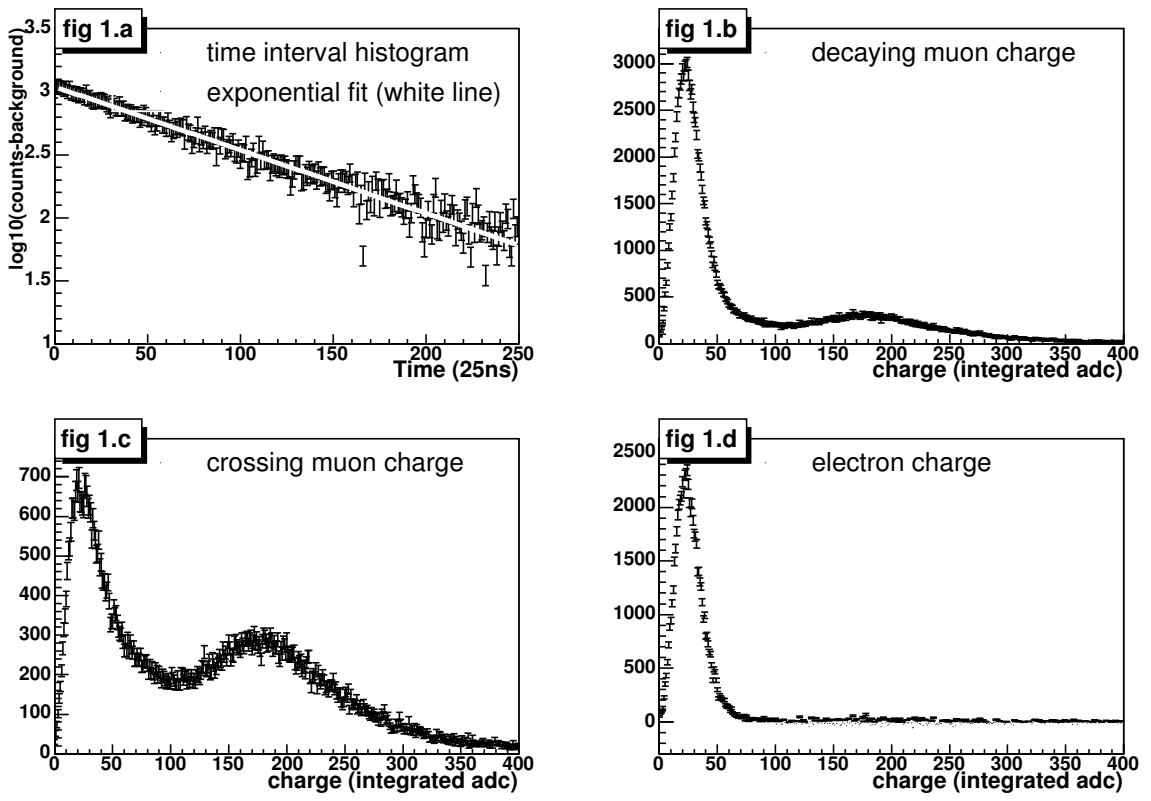

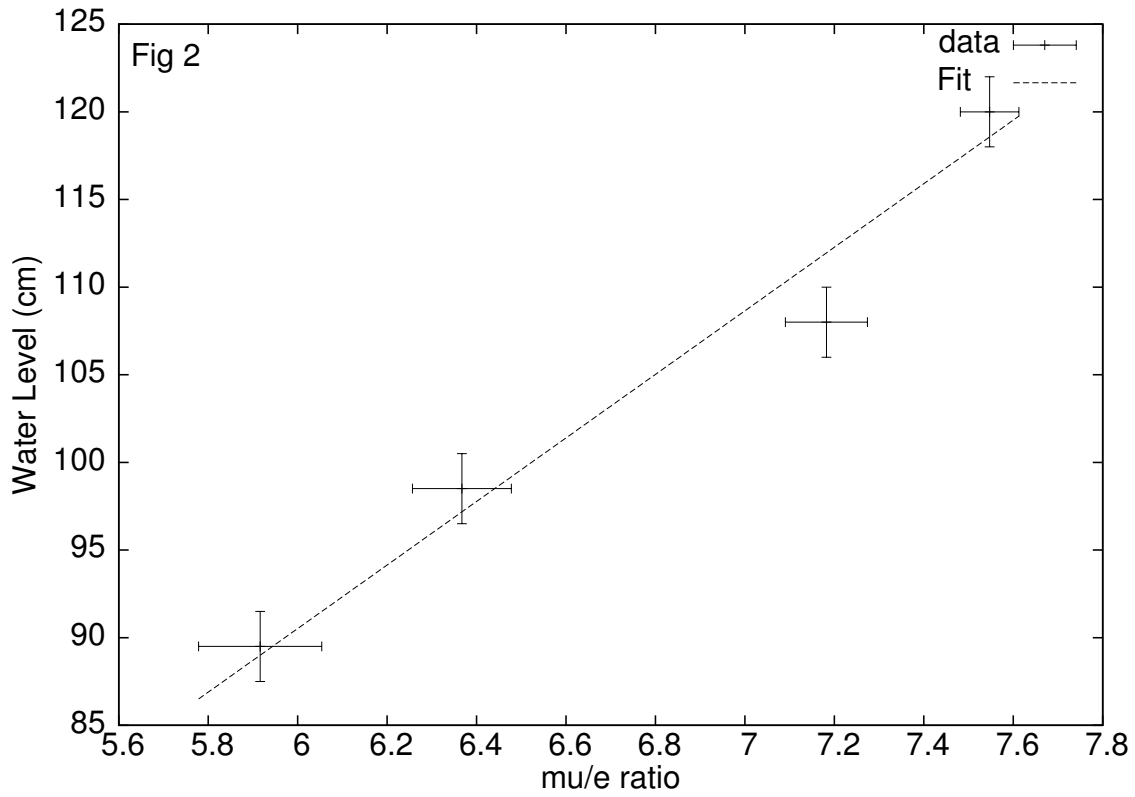

**Figure 1**(Four figures on the left). 1.a (top left) Time interval histogram and fit. To show the contribution of decaying muons, the background from crossing muons was subtracted. The time constant of the exponential is $(2.16 \pm 0.05)\ \mu s$. 1.b (top right) Charge of "early" events. 1.c (bottom left) Charge of "late" events. 1.d (bottom right) Charge deposited by a decaying muon. **Figure 2:**Figure on the right. "$\mu/e$" ratio vs water level (see text).

With this picture in mind we can obtain the charge histogram of the electrons from muon decay by subtracting the "early" events from the "late" events histogram. The resulting histogram is shown in figure 1.d. This histogram has a simple structure presenting a single peak. This peak is at a slightly different position than the first peak in figures 1.b and 1.c, implying that it is not due to the same low energy tail but a real component present in fig. 1.b and not in figure 1.c. This spectrum is the Michel spectrum convolved with the response of the detector. From Monte Carlo simulations, it can be shown that the position of the electron peak should be at $(0.124 \pm 0.006)Q_{VEM}$. We can use this value as a calibration point which, for this tank, corresponds to $(22 \pm 1)$FADC. This last value corresponds to the position of the electron peak, obtained by a gaussian fit in a reduced region around the maximum.

## 3. The Water Level Experiment.

Since the range of the electron coming from muon decay in the tank is of the order of 25 cm, most of its energy will be deposited, even if the water level of the tank is lower than the nominal value of 120 cm. On the other hand, the energy deposited by a crossing muon is proportional to its track length ($\simeq 2$ MeV cm$^{-1}$). These two magnitudes are then expected to have a different dependence on water level, while being both proportional to Tyvek reflectivity, collection efficiency, etc. (even though the energy deposited by the electron is independent of water level, a dependence of the signal produced is expected since the fraction of collection area and total area increases). It follows that the ratio of the two will depend only on water level, while other unknown factors cancel.

A simple analogy between a tank and an integrating sphere helps to understand the signal that is produced in the tank under the two circumstances referred to in the previous paragraph. From this analogy, we obtain that the signal deposited by a particle (muon or electron) in the tank is proportional to the energy deposited and a geometrical factor that depends on water level. It follows that the ratio of the signal of a crossing muon and that of an electron from a muon decay is equal to the ratio of the energy deposited by each particle, so a linear dependence is expected.

To test these assertions, the following experiment was performed: over a period of four days, we slowly drained a water tank, starting from a water level at its normal value of about 120 cm and finishing with a water level of 89.5 cm. The starting drain rate was about 1 $\ell$ every 80 s. Starting from the normal water level, the series of three histograms, output of the mufill program, were downloaded from this water tank every 24 hs. Each day the tank was visited to measure its water level. We obtained 120 cm, 108 cm, 98.5 cm and 89.5 cm for each day respectively. Figure 2 shows the water level as a function of the ratio of the muon peak to the electron peak. The error bars correspond to an error of 2 cm in the determination of the water level. The horizontal error bars are statistical.

Figure 2 also shows a linear fit to the data. From this fit we obtain the following relations between water level ($h$) and "$\mu/e$" ratio:

$$h(\mathrm{cm}) = (-18 \pm 19) + (18 \pm 3) \times \mu/\mathrm{e}$$

## 4. Results and Conclusions.

We presented a simple method to extract the charge spectrum of decaying muons in an Auger water Cherenkov detector. By comparing this charge spectrum with simulations a calibration point can be obtained, and for the tank studied the result is: $(0.124 \pm 0.006)Q_{VEM} = (22 \pm 1)$FADC or $(1.00 \pm 0.05)Q_{VEM} = (176 \pm 8)$FADC. This value is in agreement with the value of $Q_{VEM}$ obtained as explained in section 2: $1VEM = (176 \pm 1)$FADC, within experimental uncertainties.

We also presented the results of a study of the "$\mu/e$" ratio as a function of water level. An absolute determination of the water level using this method is possible, although the uncertainty is quite large as compared with the precision with which water is deployed to a tank. However, a continuous check on the online histograms could point out possible water leaks in the field. A different study using the same online histograms resulted in a more accurate procedure to monitor water level. This study is presented in [6].

## References

[1] M.Aglietta et al., Calibration of the Surface Array of the Pierre Auger Observatory, usa-allison-PS-abs1-he14-poster, these proceedings.
[2] M.Alarcon et al., Nucl.Instrum.Meth. A420:39-47, 1999
[3] C.Medina et al., http://www.auger.org/admin/GAP_Notes/Gap2003/GAP2003_091.pdf
[4] P.L.Ghia et al., http://www.auger.org/admin/GAP_Notes/Gap2003/GAP2003_070.pdf
[5] F.Arneodo et al., http://www.auger.org/admin/GAP_Notes/Gap2004/GAP2004_049.pdf
[6] I.Alekotte et al. Observation of long term stability of water tanks in the Pierre Auger surface detector, usa-arisaka-K-abs1-he15-poster, these proceedings.